# Low-temperature heat transport of spin-gapped quantum magnets

Xia Zhao[1], Zhiying Zhao[2], Xuguang Liu[2], and Xuefeng Sun[2,3,4]

[1]*School of Physical Sciences, University of Science and Technology of China, Hefei, Anhui 230026, China*

[2]*Hefei National Laboratory for Physical Sciences at Microscale, University of Science and Technology of China, Hefei, Anhui 230026, China*

[3]*Key Laboratory of Strongly-Coupled Quantum Matter Physics, Chinese Academy of Sciences, Hefei, Anhui 230026, China*

[4]*Collaborative Innovation Center of Advanced Microstructures, Nanjing, Jiangsu 210093, China*

This article reviews low-temperature heat transport studies of spin-gapped quantum magnets in the last few decades. Quantum magnets with small spins and low dimensionality exhibit a variety of novel phenomena. Among them, some systems are characteristic of having quantum-mechanism spin gap in their magnetic excitation spectra, including spin-Peierls systems, $S = 1$ Haldane chains, $S = 1/2$ spin ladders, and spin dimmers. In some particular spin-gapped systems, the XY-type antiferromagnetic state induced by magnetic field that closes the spin gap can be described as a magnon Bose-Einstein condensation (BEC). Heat transport is effective in probing the magnetic excitations and magnetic phase transitions, and has been extensively studied for the spin-gapped systems. A large and ballistic spin thermal conductivity was observed in the two-leg Heisenberg $S = 1/2$ ladder compounds. The characteristic of magnetic thermal transport of the Haldane chain systems is quite controversial on both the theoretical and experimental results. For the spin-Peierls system, the spin excitations can also act as heat carriers. In spin-dimer compounds, the magnetic excitations mainly play a role of scattering phonons. The magnetic excitations in the magnon BEC systems displayed dual roles, carrying heat or scattering phonons, in different materials.

## 1 Introduction

Quantum effects are enhanced in antiferromagnetic (AFM) systems with small spins ($S = 1/2$ or $S = 1$) and low dimensionality. The resulted quantum magnetism and quantum phase transitions (QPTs) has been a very active topic in the field of condensed-matter physics.

The most familiar model in one-dimensional quantum spins systems is the $S = 1/2$ Heisenberg model and the interaction between different spins is confined to the nearest-neighbor. The Hamiltonian can be described as

$$H = J\sum_{i=1}^{N} \vec{S}_i \cdot \vec{S}_{i+1}. \tag{1}$$

If the Hamiltonian (1) is to be treated quantum mechanically rather than classically, the ground state of $S = 1/2$ Heisenberg antiferromagnetic (HAFM) chain is a gapless spin-liquid state rather than long-range Néel state [1].

Some low-dimensional quantum spin systems have spin gap (SG). The SG can be experimentally confirmed by susceptibility, $\chi$, which behaves as $\chi \sim \exp(-\Delta/T)$ at low temperatures. The simple examples of spin-gapped antiferromagnets are the Haldane chain, spin ladder, and spin dimer. Due to the presence of the SG, magnetic properties are more complicated and peculiar than those gapless systems.

Heat transport is an efficient experimental tool in solid-state physics. It is very sensitive to even weak disorder and significantly influenced by phase transitions. In insulating materials, heat is mainly carried by phonons. However, some low-dimensional magnetic insulators have displayed the presence of another channel for heat transport, namely, a contribution of magnetic excitations.

This review is organized as follows: In Sec. 2, we give a brief overview on several kinds of typical spin-gapped systems and the origins of SG. In Sec. 3, we mainly discuss the role of magnetic excitations in the heat transport of spin-gapped systems. In Sec. 4, we discuss the heat transport in a particular class of spin-gapped systems, i.e., the magnon BEC candidates.

## 2  Spin-Gapped quantum magnets

As mentioned above, the ground state of $S = 1/2$ HAFM chain is a gapless spin liquid state. However, some other systems such as spin dimer, Haldane chain, spin ladder, and spin-Peierls systems are known to have spin gap.

### 2.1  Spin dimer

Spin-1/2 dimer is the simplest model among spin-gapped quantum magnets [2,3]. The ground state of spin dimer is spin-singlet state with total spin $S_{tot} = 0$. The wave function of spin singlet state is given by [4]

$$\Psi_s = \frac{1}{\sqrt{2}}\left(|\uparrow\downarrow\rangle - |\downarrow\uparrow\rangle\right), \quad (2)$$

where ↑ and ↓ represent the spin-up and spin-down. The energy of spin singlet state is given by

$$E_{gs} = -\frac{3}{4}J, \quad (3)$$

where $J$ is the exchange interaction between two nearest neighboring (NN) spins. The first excited state may be created by stimulating the dimer singlets to the $S = 1$ triplet state with

$$\Psi_T = \left[\frac{1}{\sqrt{2}}\left(|\uparrow\downarrow\rangle + |\downarrow\uparrow\rangle\right), |\uparrow\uparrow\rangle, |\downarrow\downarrow\rangle\right] \quad (4)$$

and

$$E_{es} = \frac{1}{4}J. \quad (5)$$

Thus, the energy gap between first excited state and ground state is $\Delta = E(S=1) - E(S=0) = J$.

### 2.2  Haldane chain

Haldane conjectured in 1983 that the HAFM spin chains with integer spin have a gap in excitation spectrum [5]. It has been verified both experimentally and theoretically later [6,7]. The HAFM spin chain with $S = 1$ is the simplest case, which is commonly known as the Haldane chain. In 1987, Affleck, Kennedy, Lieb, and Tasaki (AKLT) proposed a solvable Hamiltonian [8],

$$H_{AKLT} = \sum_{i,j} \left[ \frac{1}{2}\left(\vec{S}_i \cdot \vec{S}_j\right) + \frac{1}{6}\left(\vec{S}_i \cdot \vec{S}_j\right)^2 + \frac{1}{3} \right]. \tag{6}$$

The ground state of $H_{AKLT}$ is the valence bond solid (VBS) with a single valence bond connecting every NN pair of sites, as shown in Fig. 1. Each individual spin-1 can be considered to be a symmetric combination of two 1/2 spins [9].

$$\begin{aligned} |+\rangle &= |\uparrow\uparrow\rangle, \\ |0\rangle &= \frac{|\uparrow\downarrow\rangle + |\downarrow\uparrow\rangle}{\sqrt{2}}, \\ |-\rangle &= |\downarrow\downarrow\rangle. \end{aligned} \tag{7}$$

On the NN sites, the pairs of spin-1/2 are linked in a singlet state [9]:

$$\frac{|\uparrow\downarrow\rangle - |\downarrow\uparrow\rangle}{\sqrt{2}}. \tag{8}$$

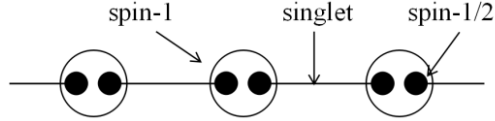

**Figure 1**  The ground state of AKLT Hamiltonian [9].

The excitation spectrum of $H_{AKLT}$ is very difficult to determine exactly. But a lot of researches have shown that there is a spin gap in the excitation spectrum [10,11]. This result is now well supported by numerical calculations on spin-1 1D HAFM ring. A reliable estimation of the energy gap, $\Delta = 0.41|J|$ at the wave vector $q = \pi$ in the excitation spectrum, where $J$ is the NN exchange, has been obtained from Monte Carlo calculations up to 32 spins [12].

## 2.3  Spin ladder

The simplest spin ladder model composes of two spin-1/2 chains coupled through the rung exchange interaction ($J_R$) and the NN intra-chain exchange ($J$) (Fig. 2). It is a spin system situated between one-dimensional (1D) and two-dimensional (2D) quantum magnets [13-15].

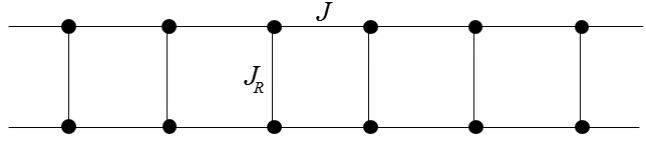

**Figure 2**  A two-chain spin ladder.

For all $J_R/J > 0$, there is a gap in the spin excitation spectrum [16]. It can be easy to understand in the simple limitation of $J_R$ much stronger than $J$. The $J$ may be treated as a perturbation under this circumstance. When $J = 0$, the exact ground state is composed by singlets along the rungs and the energy of ground state is $-(3J_R N/4)$, where $N$ is the number of rungs. Obviously, the total spin of ground state is zero. The lowest excited state in the ladder may be created by exciting one of the rung singlets to the $S = 1$ triplet state. It should be noted that the weak NN intra-chain exchange interaction gives rise to propagating $S = 1$ magnons. The dispersion relation of $S = 1$ magnons can be described by

$$\omega(k) = J_R + J \cos k ,  \qquad (9)$$

where $k$ is the momentum wave vector. The spin gap is given by

$$\Delta = \omega(\pi) \approx J_R - J . \qquad (10)$$

The spin-spin correlation decays exponentially along the chain, indicating that the ground state is a quantum spin liquid [16].

In general, if there are $n$ spin chains in spin ladder system, the excitation spectrum is gapped (gapless) when $n$ is even (odd) [17].

## 2.4  Spin-Peierls system

The Spin-Peierls (SP) system is historically the first example of magnetic systems exhibiting SG [18]. The SP transition was originally observed in some organic compounds [19], and was found in an inorganic compound, $CuGeO_3$, for the first time by Hase *et al.* in 1993 [20]. For the occurrence of a SP transition, several preconditions are necessary: firstly, a crystal must contain (quasi-) 1D AFM chains of half-integer spin; secondly, the spin-phonon coupling is strong. The magnetic energy can be lowered by a SP transition, which may simply be described as follows: below a certain transition temperature, $T_{SP}$, the distances between neighboring spins are no longer uniform. Due to the spin-phonon coupling, an alternation of the exchange coupling and each pair of coupled spins is forming a spin singlet. This spin dimerization leads to a gain of magnetic energy which over-compensates the loss of elastic energy arising from the alternating structural distortion along the spin chains. Below $T_{SP}$, a periodic deformation of the lattice sets in. In this sense, the SP transition is a three-dimensional (3D) structural phase transition that is driven by the 1D magnetism.

## 2.5  Magnon Bose-Einstein condensation system

Bose-Einstein condensation (BEC) represents a collective occupation of bosons to the lowest single-particle state at temperatures very close to the absolute zero. For some particular spin-gapped quantum magnets, the XY-type AFM state induced by the external magnetic field that close the spin gap can be described as a magnon BEC state [21-29]. It can result in a QPT from the low-field disordered phase to the high-field long-range ordered state. The theoretical research on magnon BEC could date back to 1956 by Matsubara and Matsuda [30], while $TlCuCl_3$ was the first experimentally realized material [31]. As a peculiar quantum state, magnon BEC has been extensively studied for many spin-gapped quantum magnets [29].

## 3 Heat transport in spin-gapped systems

### 3.1 Heat transport by magnetic excitations

Magnetic excitations participating heat transport was firstly predicted in 1936 [32] and confirmed in ferrimagnet YIG after about thirty years [33-35]. In addition to contributing to heat transport, the magnetic excitations can also scatter phonons and reduce the phonon thermal conductivity. In general, magnetic excitations either transport heat or scatter other quasi-particles, or scattering effect and heat transport exist simultaneously. And the overall effect depends on the relative strength of the two mechanisms in the different temperature regions.

The energy transport in 1D spin systems has been addressed employing either the linear response formalism or the Boltzmann transport equation formalism. The latter approach, which relies on the quasi-particle picture with associated velocities and relaxation times, provides transparent results and is well applicable for the analysis of experimental data. However, this quasi-particle mode is not always suited for quantum many-body systems. Thus, the most widely used approach was directed toward the calculation of transport coefficients via time-dependent current-current correlation functions, the thermal conductivity at a finite frequency is [36-38]:

$$\kappa(\omega) = \frac{1}{T} \int_0^\infty dt e^{-i\omega t} \int_0^{1/T} d\tau \left\langle j_{th}(-t-i\tau) j_{th} \right\rangle, \quad (11)$$

where <...> denotes the thermodynamic average and $j_{th}$ is the energy current. The real part of the formula can be decomposed into [36-38]:

$$Re\kappa(\omega) = D_{th}\delta(\omega) + \kappa_{reg}(\omega), \quad (12)$$

where the weight of the singular part ($D_{th}$) is the so-called thermal Drude weight, which implies that energy current is ballistic or diffusive.

Theories and experiments have already proved that the $S = 1/2$ Heisenberg chain system, an integrable spin model, exhibits a ballistic transport of the magnetic excitations [38-42]. However, the behaviors of heat transport of quantum spin-gapped systems are much more complicated.

### 3.2 Heat transport in $S = 1/2$ spin ladder

The typical two-leg $S = 1/2$ spin-ladder material, $(Sr,Ca,La)_{14}Cu_{24}O_{41}$, has been found to display a ballistic spin-mediated transport [43-48]. $(Sr,Ca,La)_{14}Cu_{24}O_{41}$ is often called telephone-number compound because of the six digits in the chemical formula. Both spin ladders and chains extend in the $c$ direction. A large spin gap in the ladders of the order of 400 K was observed in this system [14,49-51].

Figure 3 shows the thermal conductivity of $Sr_{14}Cu_{24}O_{41}$ as a function of temperature measured along the three crystallographic axes [44]. A large anisotropy can be observed both in magnitude and temperature dependence of the $\kappa$. The most notable phenomenon is the behavior of $\kappa_c$ at $T > 50$ K, where a second pronounced maximum of the $\kappa$ occurs at $T \approx 140$ K. It is reasonable to attribute the high-$T$ maximum of $\kappa_c$ to magnetic excitations propagating along the spin ladders. In order to estimate the phonon background, a Debye model was used to fit the data [43,44],

$$\kappa_{ph} = \frac{k_B}{2\pi^2 v}\left(\frac{k_B}{\hbar}\right)^3 \int_0^{\Theta_D} \frac{x^4 e^x}{(e^x-1)^2}\tau(\omega,T)dx, \quad (13)$$

in which $x = \hbar\omega/k_B T$, $\omega$ is the phonon frequency, and $\tau(\omega,T)$ is the mean lifetime of phonon. The phonon relaxation rate is usually defined as

$$\tau^{-1}(\omega,T) = v/L + A\omega^4 + BT\omega^3 \exp\left(-\frac{\Theta_D}{bT}\right) + \tau_{res}^{-1}, \quad (14)$$

where the four terms correspond to phonon scattering by the grain boundaries, the point defects, the phonon-phonon $U$ processes, and the resonant scattering, respectively. $L$, $A$, $B$, and $b$ are free parameters. The result of this fitting of $\kappa_{ph}$ is shown by a solid line in Fig. 3, which is much smaller than the measured $\kappa_c$ in $T > 50$ K. The significant deviation of $\kappa_c$ from the Debye fitting at $T > 50$ K is due to the large magnetic heat transport.

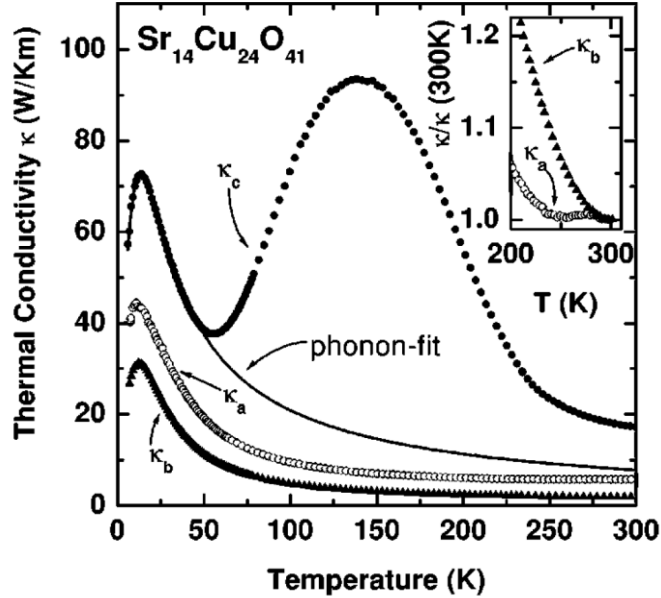

**Figure 3** Thermal conductivity of $Sr_{14}Cu_{24}O_{41}$ measured parallel ($\kappa_c$) and perpendicular to the ladders and chains ($\kappa_a$, $\kappa_b$) [44]. The solid line is a fit to the low-temperature peak to describe the phonon contribution to $\kappa_c$. Inset: $\kappa_a$ and $\kappa_b$ normalized to the value at 300 K.

Thermal conductivities of $Ca_9La_5Cu_{24}O_{41}$ along the $a$ and $c$ axes are shown in Fig. 4 [44]. Similar to $Sr_{14}Cu_{24}O_{41}$, $\kappa_c$ strongly increases at $T > 40$ K, and is even larger than of $Sr_{14}Cu_{24}O_{41}$ at intermediate and high temperatures. In contrast, a very small thermal conductivity was found along the $a$ axis in the entire temperature range. As shown in Fig. 4, it is obvious the $\kappa_{ph}$ can be inferred from $\kappa_a$ as well as from the low-$T$ behavior of $\kappa_c$.

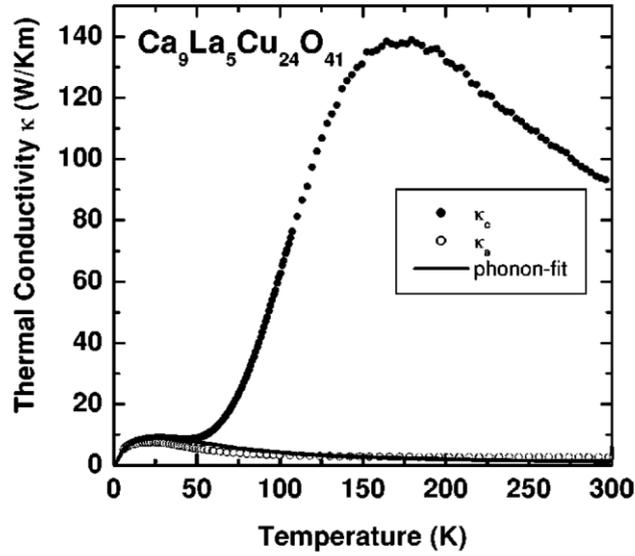

**Figure 4** Thermal conductivities of $Ca_9La_5Cu_{24}O_{41}$ as a function of temperature measured along the $a$ and $c$ axes, $\kappa_a$ and $\kappa_c$, respectively [44]. The solid line represents an estimate of the phonon contribution to $\kappa_c$.

Figure 5 shows the $\kappa_{mag}$ for $Sr_{14}Cu_{24}O_{41}$ and $Ca_9La_5Cu_{24}O_{41}$, which are derived by subtracting the Debye fits of the phonon contribution from the measured $\kappa_c$ [44]. For $T < 100$ K, the magnitudes and temperature dependencies of $\kappa_{mag}$ are similar in the two compounds and a large magnon heat transport was found. However, pronounced differences occur at higher temperatures. The magnetic heat conductivity of $Sr_{14}Cu_{24}O_{41}$ strongly decreases above 150 K. This decrease is much less pronounced and is found at higher temperature above 200 K in $Ca_9La_5Cu_{24}O_{41}$, which has a very large $\kappa_{mag}$ of 100 W/Km even at room temperature.

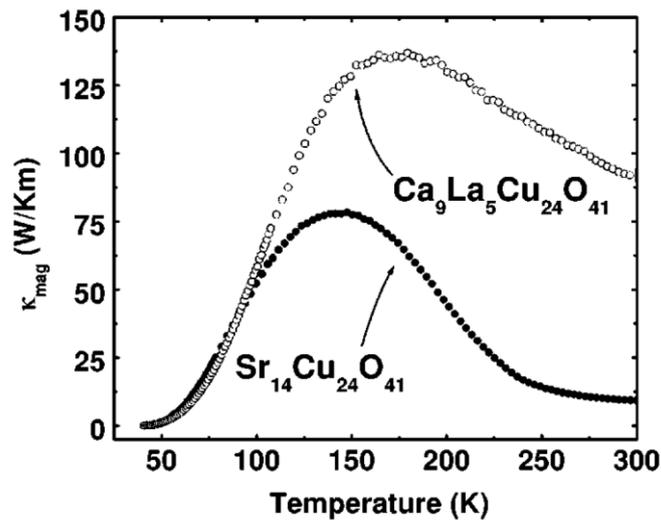

**Figure 5** Magnon thermal conductivity of $Sr_{14}Cu_{24}O_{41}$ and $Ca_9La_5Cu_{24}O_{41}$, obtained by subtracting the phonon contribution from the raw data [44].

A simple kinetic was applied to describe this magnetic contribution,

$$\kappa_{mag} = \frac{3Nl_{mag}}{\pi\hbar k_B T^2} \int_{\Delta_{ladder}}^{\varepsilon_{max}} \frac{\exp(\varepsilon/k_B T)}{[\exp(\varepsilon/k_B T)+3]^2} \varepsilon^2 d\varepsilon . \tag{15}$$

Here, $N$ is number of ladders per unit area and $\varepsilon_{max}$ is the band maximum of the spin excitations, which is approximately 200 meV in $(Sr,Ca,La)_{14}Cu_{24}O_{41}$. From calculations of $\kappa_{mag}$, the spin gaps are $\Delta_{ladder}=$ 396 K and $\Delta_{ladder}=$ 430 K for $Ca_9La_5Cu_{24}O_{41}$ and $Sr_{14}Cu_{24}O_{41}$, respectively, which are in fair agreement with the results from neutron scattering [14]. Based on these analyses of the low-$T$ behavior of $\kappa_{mag}$, the authors could determine the temperature dependence of the mean free path of magnons, which showed a very large value of several thousand Å at low temperatures [44].

Another interesting way to study the spin gap properties in spin ladder is to replace Sr by isovalent Ca which initiates a transfer of holes from the chains to the ladders, leading to a change of the temperature dependence of the $c$-axis resistivity from semiconducting to metallic [43]. The substitution does not change $\Delta_{chain}$, indicated by inelastic neutron scattering [51], but the ordered state of dimers in the chains becomes unstable.

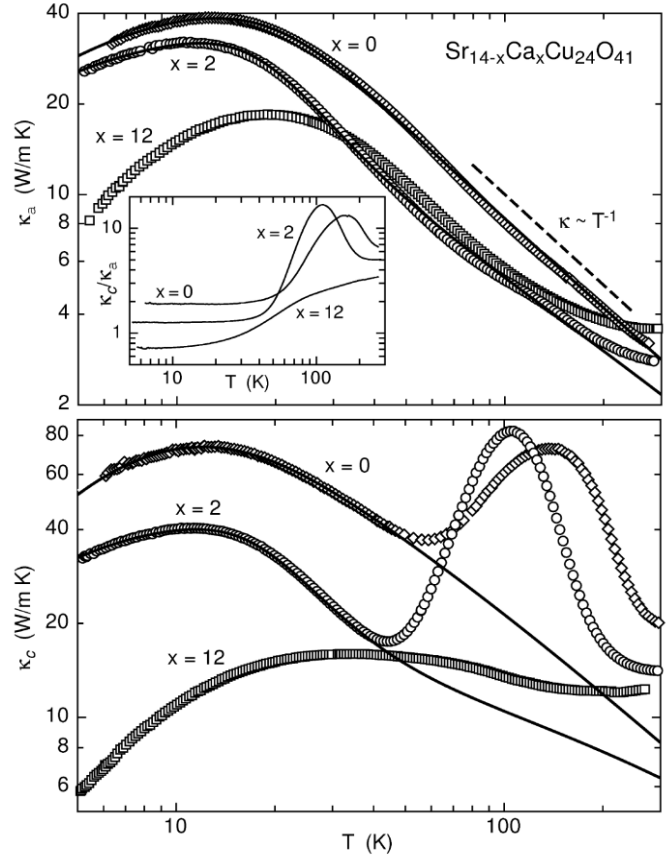

**Figure 6**  Temperature dependence of the thermal conductivity of $Sr_{14-x}Ca_xCu_{24}O_{41}$ ($x$ = 0, 2, and 12) along the $a$ ($\kappa_a$) and the $c$ axis ($\kappa_c$) [43]. The solid lines represent the calculated phonon thermal conductivities. The inset shows the temperature dependences of anisotropy ratio $\kappa_c/\kappa_a$.

Figure 6 shows the temperature dependence of the thermal conductivity of $Sr_{14-x}Ca_xCu_{24}O_{41}$ ($x = 0$, 2, and 12) along the $a$ and the $c$ axis [43]. A low-$T$ peak can be seen in $\kappa_a$, and at high temperatures, the temperature dependence is close to $\kappa_a \sim T^{-1}$, which is typical for phonon heat transport. The electronic thermal conductivities were found to be negligible for semiconducting $x = 0$ and 2 samples and less than 3% of the total $\kappa$ for $x = 12$. At temperatures below 30 K, small anisotropy ratio $\kappa_c/\kappa_a$ was observed, as shown in the inset to Fig. 6. It indicated that the $\kappa_c$ is predominantly phononic and has the same types of phonon scattering to determine the behavior. However, for temperature above 40 K, $\kappa_c(T)$ is qualitatively different from $\kappa_a(T)$, especially for $x = 0$ and 2. And $\kappa_c(T)$ reveals an excess thermal conduction exhibiting peak values above 100 K.

The experimental data of $\kappa_a(T)$ at all temperatures and of $\kappa_c(T)$ below 30 K were fitted to the formula (13) given by the Debye model. And the extrapolated $\kappa_{c,ph}$ (see Fig. 6) has been subtracted from the total measured $\kappa_c$. The data are plotted on a logarithmic scale versus $1/T$ in the inset to Fig. 7 [43]. It can be seen that between 40 and 65 K the excess thermal conductivities vary as $\exp(-\Delta_\kappa/T)$ with $\Delta_\kappa = 355 \pm 40$ and $363 \pm 40$ K for $x = 0$ and 2, respectively. Because $\Delta_\kappa$ coincides with $\Delta_{ladder}$, the data provided additional support for the thermal transport by magnetic excitations. In addition, an evident magnetic thermal transport was also observed in the $x = 12$ samples but strongly damped, probably due to the scattering effect of the enhanced number of itinerant holes.

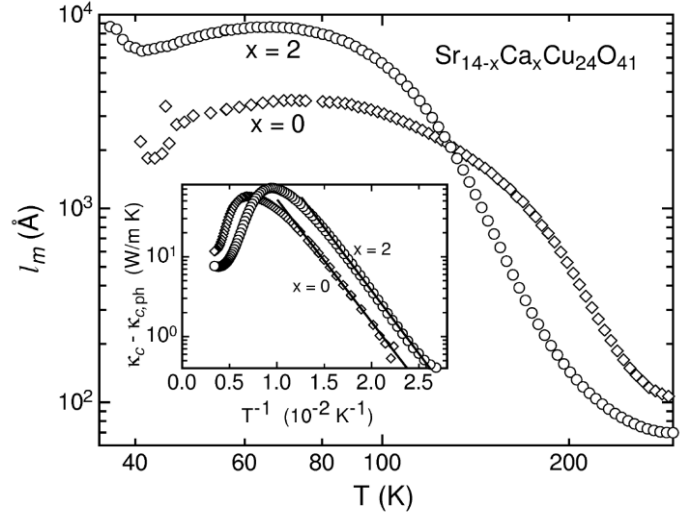

**Figure 7** Mean free path of spin excitations for $Sr_{14-x}Ca_xCu_{24}O_{41}$ ($x = 0$ and 2) [43]. The inset is ($\kappa_c - \kappa_{c,ph}$) vs $T^{-1}$; the solid lines represent an exponential temperature variation of the type $\exp(-\Delta_\kappa/T)$.

Mean free path of spin excitations $l_m$ can be obtained from the fittings and is plotted in Figure 7. For $x = 0$ and 2, $l_m$ is weakly temperature dependent up to 80–100 K, then decreases rapidly and tends to saturate at $T \geq 250$ K. Since a triplet excitation propagating along the ladder can pass through a hole pair, the decreasing number of unpaired holes below 250 K should enhance the mean free path of magnetic excitations, in agreement with the data shown in Fig. 7. And the almost constant magnon mean free path at $T < 80$ K is probably due to crystal defects and remaining unpaired holes.

The experimental results all indicated a ballistic thermal transport of magnetic excitations in the $S = 1/2$ ladder system, $(Sr,Ca,La)_{14}Cu_{24}O_{41}$. However, theoretical studies on the thermal conductivity of spin-1/2 ladders are rather controversial [52-57]. On the one hand, the spin-1/2 ladder is believed to have the essentially same as the spin-1 chain in the aspects of the spin-liquid ground state and the gapped magnetic spectrum. The pioneer theoretical result indicated a diffusive spin transport of the spin-1 chain system, which is a non-integrable model [52]. On the other hand, some recent theories also suggested a more complicated spin transport behavior in spin-1/2 ladders [53-57]. It is possible that this system could exhibit large bust diffusive spin thermal conductivity. Particularly, Steinigeweg *et al.* demonstrated that the inter-chain coupling is an important parameter determining the thermal conductivity of the Heisenberg spin-1/2 ladder [57].

### 3.3 Heat transport in $S = 1$ Haldane chain

As more and more Haldane chain materials have been found, the heat transport experiments of these materials were widely carried out. Sologubenko *et al.* have studied heat transport of the typical Haldane material $AgVP_2S_6$ [58]. This material has the intra-chain exchange constant $J/k_B = 780$ K and a weak single-ion anisotropy $D$ ($D/J = 5.83 \times 10^{-3}$). The spin gap $\Delta$ is roughly equal to 300 K [59,60].

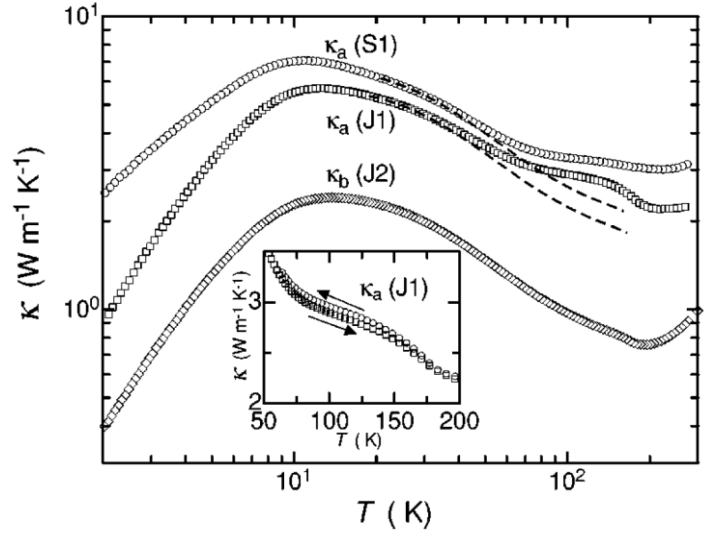

**Figure 8** Thermal conductivity of the $AgVP_2S_6$ for two samples (S, J); measured parallel ($\kappa_a$) and perpendicular ($\kappa_b$) to the chains [58]. The inset shows the hysteresis in $\kappa_a$ at high temperatures.

The thermal conductivities of $AgVP_2S_6$ for two samples are shown in Fig. 8 [58]. All curves exhibit the same qualitative feature below 60 K. However, at higher temperatures, a shoulder-type feature in $\kappa_a(T)$, which is absent in the $\kappa_b(T)$ curve, can be identified. Since $\kappa_b(T)$ is totally phononic, an obvious interpretation of this high-$T$ feature is to ascribe it to a spin-mediated thermal conductivity $\kappa_m$ in the $a$ direction (spin-chain direction).

Assuming that at high temperatures, the phonon thermal conductivity along different directions and for different samples merely differs by a constant factor, the magnon contribution can be calculated as

$$\kappa_{m,a} = \kappa_a - K\kappa_b, \qquad (16)$$

with the coefficients $K = 2.7$ and 2.3 for samples S1 and J1, respectively. The phonon contributions $K\kappa_b$ for samples S1 and J1 are shown by the broken lines in Fig. 8.

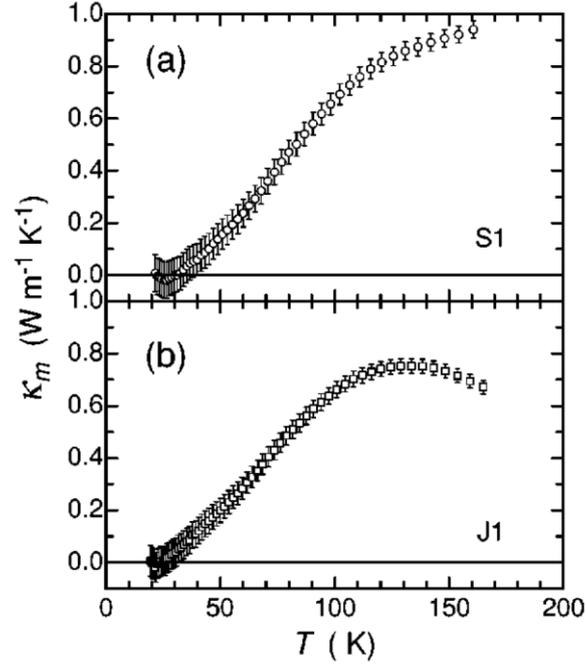

**Figure 9**  Magnon thermal conductivity along the *a* axis of AgVP$_2$S$_6$ [58].

The obtained values of $\kappa_m(T)$ are presented in Fig. 9. Sologubenko *et al.* mainly focused on an analysis of the energy diffusion constant [58]:

$$D_E = \kappa_m / \left[ C_s(T) a^2 \right]. \tag{17}$$

$C_s(T)$ is the specific heat of the spin system and $a = 2.96$ Å is the distance between neighboring spins along the chains. In comparison with the spin diffusion constant $D_s$, which is experimentally known for this compound from Ref. [61], these two quantities turn out to be of the same order of magnitude with $D_E \approx 2D_s$. An additional analysis of the magnetic mean free paths resulted in very small values ($\leq 60$ Å) and led to the conclusion of diffusive transport in this Haldane-chain compound.

The heat transport of another Haldane chain compound Y$_2$BaNiO$_5$ was studied by Kordonis *et al.* [62]. This compound has the intra-chain exchange constant $J/k_B = 250–280$ K, the gap $\Delta \approx 100$K, and $J_\perp/J < 10^{-4}$ ($J_\perp$ is the inter-chain coupling) [63-66]. Similar to AgVP$_2$S$_6$, a shoulder-like feature between about 80 and 140 K is observed only for $\kappa_a$, as shown in Fig. 10 [62]. The explanation is the existence of an extra contribution to the thermal conductivity, beside the phonon conductivity ($\kappa_{ph}$), coming from spin excitations ($\kappa_s$). The spin thermal conductivity along the chains $\kappa_s(T) = \kappa_a(T) - \kappa_{ph}(T)$ is plotted in the inset of Fig. 10(a). The energy diffusion constant $D_E(T)$ could be calculated using formula (17).

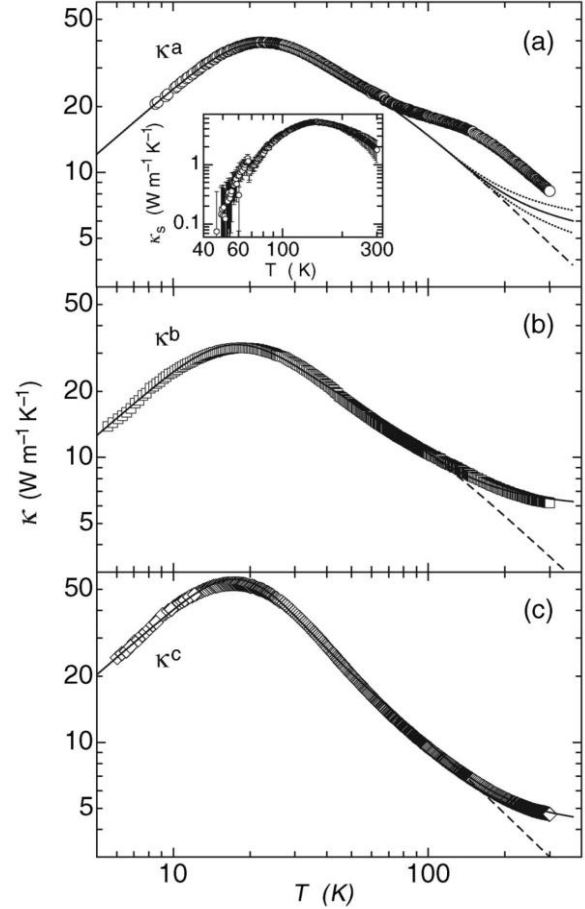

**Figure 10** Thermal conductivity of $Y_2BaNiO_5$ along (*a* axis) and perpendicular (*b* and *c* axis) to the spin chains [62]. The solid (dashed) lines represent calculations of the phonon contributions. The inset shows the spin contribution to the thermal conductivity along the chain direction.

The results of the $Y_2BaNiO_5$ together with the data for $AgVP_2S_6$ are shown in Fig. 11 [62]. It is obvious that not only the magnitudes, but also their temperature dependencies are similar. The energy diffusion constant $D_E(T)$ remains rather low in the whole investigated temperature region for both materials and was found to be close to the predicted high-$T$ limit. This is in striking contrast to the $S = 1/2$ chains, where $D_E(T)$ increases by about 2 orders of magnitude with decreasing temperature, as shown in Fig. 11.

The above results for $Y_2BaNiO_5$ and $AgVP_2S_6$ indicated diffusive transport of magnetic excitations in the Haldane systems. However, in an organic $S = 1$ Haldane chain compound $Ni(C_2H_8N_2)_2NO_2ClO_4$ (NENP), which has relatively weaker spin interaction and smaller spin gap (12.2 K), the magnetic heat transport was found to be rather large [67]. NENP has intra-chain exchange constant $J = 43$ K and $J'/J = 8 \times 10^{-4}$ with chains along the *b* axis ($J'$ is the inter-chain interaction) [68,69]. In NENP, the strong planar anisotropy and weak orthorhombic anisotropy with $D/J = 0.2$ and $E/J = 0.01$ split $\Delta$ into three gaps $\Delta_1 = 29$ K, $\Delta_2 = 14.3$ K, and $\Delta_3 = 12.2$ K [69]. With increasing field ($B \parallel b$), $\Delta_1$ stays constant, $\Delta_2$ increases, and $\Delta_3$ decreases such that it should close at the critical field $B_c \approx 10$ T [70]. However, due to staggered transverse field which is proportional to the homogeneous field $B$, the gap remains finite at $B = B_c$ and increases above $B_c$ [71].

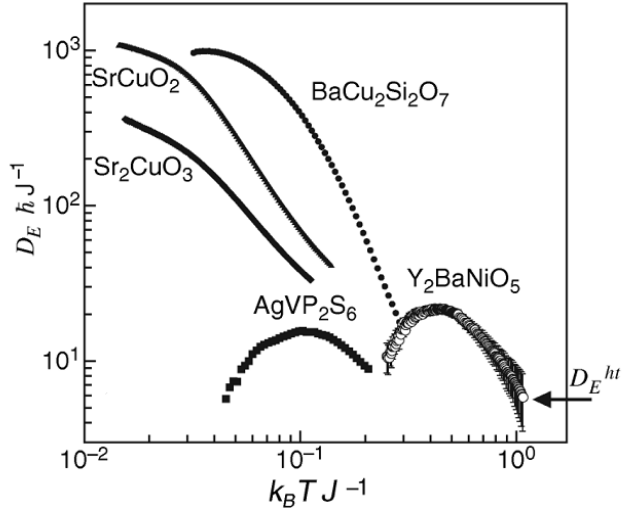

**Figure 11** The energy diffusion constant $D_E(T)$ calculated from the thermal conductivity data of $Y_2BaNiO_5$ [62]. For comparison, $D_E(T)$ of the $S = 1$ chain compound $AgVP_2S_6$ and of the $S = 1/2$ chain compounds $BaCu_2Si_2O_7$, $Sr_2CuO_3$, and $SrCuO_2$ are also shown.

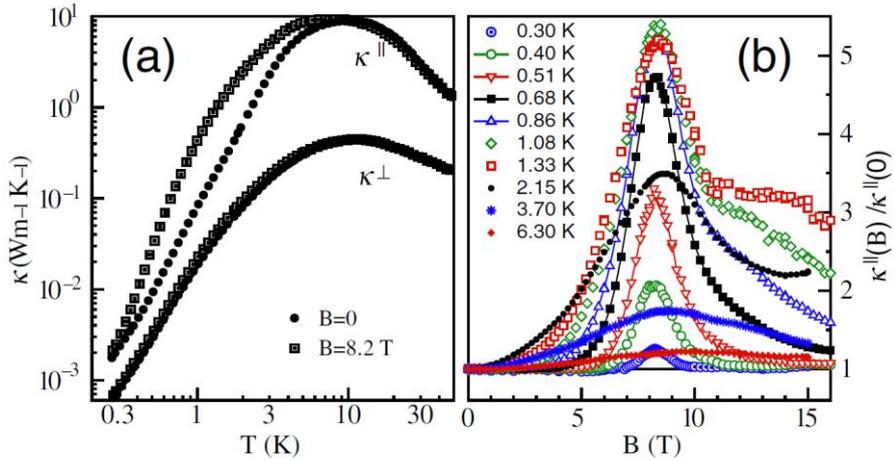

**Figure 12** (Color online) (a) Thermal conductivity of NENP parallel and perpendicular to the spin chains as a function of temperature at $B = 0$ and $B = 8.2$ T. (b) The relative change of thermal conductivity of NENP parallel to the spin chains as a function of magnetic field [67].

Figure 12(a) shows the thermal conductivity parallel ($\kappa^\parallel$) and perpendicular ($\kappa^\perp$) to the chain direction in zero field and in $B = 8.2$ T [67]. Obviously, the magnetic field leads to a strong enhancement of the thermal conductivity. The relative changes of $\kappa^\parallel$ as a function of magnetic field at several constant temperatures are shown in Fig. 12(b). Thermal conductivity in magnetic field is nearly 5 times its zero-field value at $T = 1.08$ K. This strong enhancement was not observed for $\kappa^\perp$. The magnetic thermal conductivity $\kappa_s(B,T) = \kappa^\parallel(B,T) - \kappa^\parallel(0,T)$ was obtained and is shown in Fig. 13. Based the calculation of magnetic thermal conductivity using formula (15), the mean free path of spin excitations could be obtained [67]. It was found that the mean free path is independent of both $T$ and $B$, with the average value $l_m = 0.75 \pm 0.1$ μm. Remarkably, it is as large as the highest values of the mean free path found for the $S = 1/2$ Heisenberg chains and ladders [40,43]. The absence of a temperature and field dependence of the mean free path can be explained by rare defects, which cut the spin chains into segments, in combination with a tiny inter-chain coupling.

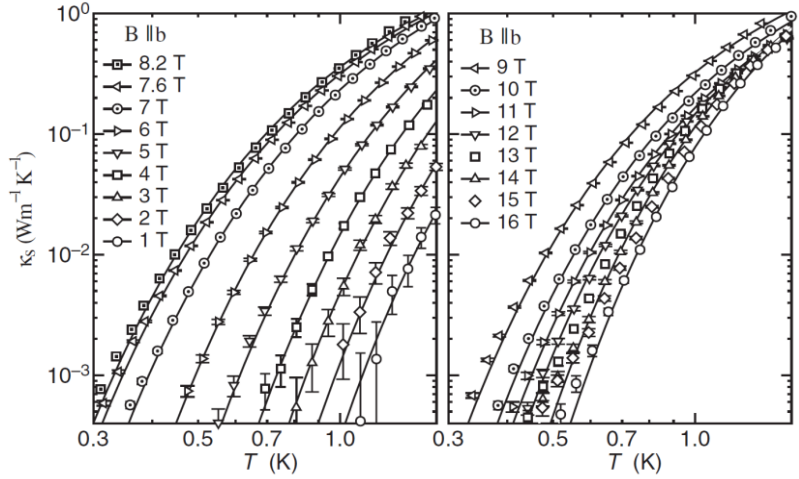

**Figure 13** Magnetic contribution to the thermal conductivity of NENP parallel to the spin chains as a function of temperature at several constant fields [67]. The solid lines represent the calculated $\kappa_s$.

It should be noted that the spin transport of NENP can only be observed in magnetic field [67], which weakens the spin gap. It is possible that the large spin transport in zero magnetic field can be found in $S = 1$ chain systems with smaller energy gaps, for example, $Ni(C_3H_{10}N_2)_2NO_2ClO_4$ (NINO) [72]. NINO has a similar spin structure to that of NENP (see Fig. 14) and was found to be another ideal $S = 1$ Haldane chain system [71]. The $Ni^{2+}$ spins ($S = 1$) form the spin chains along the $b$ axis, in which the intra-chain AFM interaction ($J = 50$ K) is about a factor of $10^4$ times stronger than the inter-chain interaction ($J'$) [68,73,74]. It is known that in an isotropic AFM $S = 1$ chain system, the spin excitations are triply degenerate with an energy gap $E_g = 0.41J$, which is about 20.3 K for NINO [75]. However, due to the strong planar anisotropy and weak orthorhombic anisotropy, the Haldane gap is split into three gaps $\Delta_1 = 8.3$ K, $\Delta_2 = 12.5$ K, and $\Delta_3 = 21.9$ K. When a magnetic field is applied along the $a$ axis, $\Delta_2$ keeps constant, $\Delta_3$ increases, and $\Delta_1$ decreases. The smallest gap $\Delta_1$ is apparently the most important for the low-energy magnetic excitations. In particular, $\Delta_1$ descends to a small value at a critical field $H_c \approx 9$ T, and then increases above $H_c$ [68,74]. In general, magnetic properties of NINO are very similar to those of NENP, except that the energy scales of the spin gap are different.

Figure 14(c) shows the temperature dependencies of $\kappa_b$ and $\kappa_c$ of NINO single crystals in zero magnetic field [72]. Apparently, the behaviors of $\kappa_b$ and $\kappa_c$ seem to be different from usual phonon transport properties of insulators [76]. In nonmagnetic insulators, the position and the magnitude of the low-$T$ phonon peak are mainly determined by the impurity scattering. Therefore, the phonon peaks are usually located at similar temperatures for different directions of heat currents and the shapes (temperature dependencies) of peaks are almost identical for different directions. These are clearly different to what the NINO data show. The most remarkable phenomenon in Fig. 14(c) is that the $\kappa_b$ and $\kappa_c$ differ significantly at relatively high temperatures, but they are almost the same at subKelvin temperatures. All these features suggest that the heat transport of NINO cannot be a simply phononic behavior. One may naively expect that the difference is due to the additional magnetic thermal transport along the chain direction.

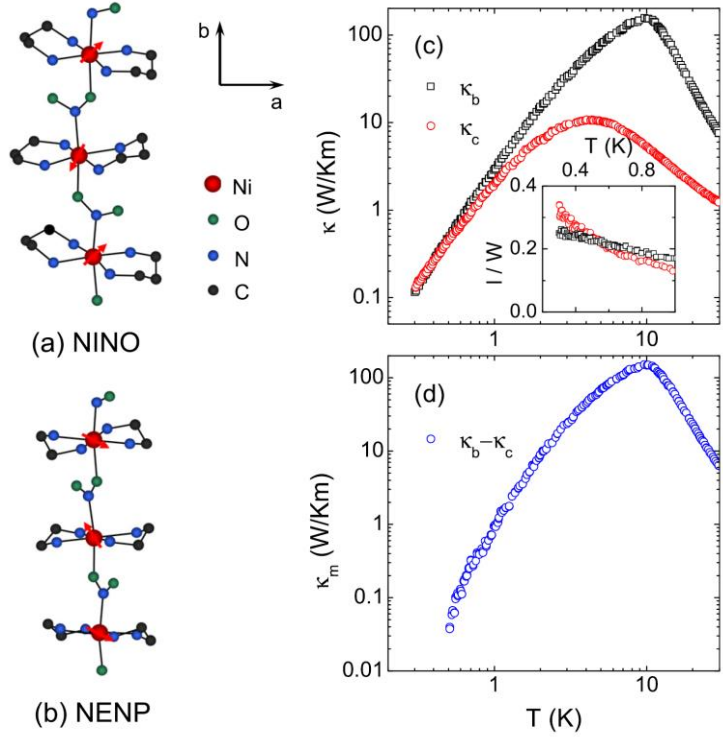

**Figure 14** (Color online) (a) and (b) Schematic view of the spin-chain structure of NINO and NENP. (c) Temperature dependencies of the thermal conductivities $\kappa_b$ and $\kappa_c$ of NINO single crystals in zero magnetic field. The inset shows the temperature dependencies of the ratio of the phonon mean free path $l$ to the averaged sample width $W$. (d) Magnetic thermal conductivity $\kappa_m(T)$ obtained from $\kappa_b-\kappa_c$ [72].

It is useful to estimate the mean free path of phonons, $l$, by using the kinetic formula $\kappa_{ph} = 1/3Cv_p l$, where $C = \beta T^3$ is the phonon specific heat at low temperatures and $v_p$ is the averaged sound velocity [77,78]. The coefficient $\beta$ is known from the specific heat measurements [79], and $v_p$ can be obtained from $\beta$. The inset to Fig. 14(c) shows the temperature dependencies of the ratio $l/W$ for $\kappa_b$ and $\kappa_c$, where $W$ is the averaged sample width. It is found that in both cases the $l$ becomes comparable to $W$ at 0.3 K, indicating that the phonon heat transport approaches the boundary scattering limit [77,78]. With increasing temperature, the $\kappa_b$ and $\kappa_c$ show a large difference because the magnetic excitations become populated and they can either transport heat along the spin-chain direction or scatter phonons in other directions.

The magnetic thermal conductivity along the spin chain can be obtained by subtracting the $\kappa_c$ from $\kappa_b$ [72]. As shown in Fig. 14(d), the magnetic thermal conductivity is very large and reaches a value of ~ 150 W/Km at 10 K. Thus, NINO has larger magnetic thermal conductivity than many other low-dimensional quantum magnets [58,62,67].

A lot of theoretical researches on heat transport of Haldane chain systems have been performed, but the results of the magnetic heat transport are rather controversial. Orignac *et al.* obtained the energy current from the nonlinear model that describes integer spin chains in the limit $S\rightarrow\infty$ [80]. The evidenced translation invariance implies that the total thermal current is conserved, which means that the magnetic heat transport is ballistic. Non-diffusive magnetic transport of Haldane chains at finite temperatures was also evidenced by Fujimoto upon the Bethe ansatz exact solution for the $O(3)$ sigma model and the $1/N$-expansion approach for the $O(N)$ sigma model [61]. However, Karadamoglou and Zotos used the exact diagonalization and microcanonical Lanczos method to study the spin thermal conductivity and obtained the temperature dependencies of thermal Drude weight under the framework of linear-response theory, which shows a diffusive magnetic heat transport of Haldane chain system [52].

### 3.4 Heat transport in the spin dimer system

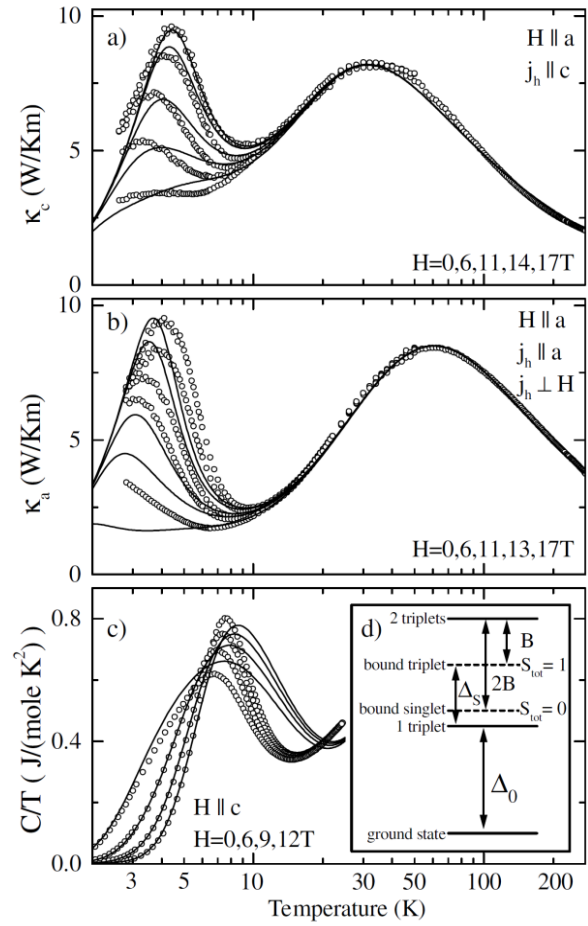

**Figure 15** Thermal conductivity $\kappa_c$ (a) and $\kappa_a$ (b) of SrCu$_2$(BO$_3$)$_2$ on a logarithmic temperature scale for various magnetic fields. On increasing field the low-$T$ maximum is suppressed. (c) Specific heat at various fields for $H \parallel c$. The lines are theoretical curves calculated for the same magnetic fields using the Debye model for the phononic thermal conductivity with considering the resonant scattering of phonons on magnetic excitations. (d) Level scheme: The solid lines denote the energies of the ground state, one triplet and two elementary triplets. The dashed lines indicate bound states with $S_{tot} = 0$ or 1, which are built from two elementary triplets. $B$ is the binding energy and $\Delta_0$ is the elementary gap. $\Delta_s$ is the energy difference relevant to the resonance scattering [81].

The thermal conductivity of SrCu$_2$(BO$_3$)$_2$ as a function of temperature and magnetic field were presented by Hofmann *et al.* [81] SrCu$_2$(BO$_3$)$_2$ can be considered as a 2D spin-gap system with orthogonal dimer network and the well-known Shastry-Sutherland model [82-84]. The intra-dimer and inter-dimer couplings are of magnitude $J_1 \approx 72$ K and $J_2 \approx 43$ K, i.e., $J_2/J_1 \approx 0.6$ [85]. The unique spin arrangement in this compound has led to extensive interest in studying this low-dimensional quantum spin system, where the ground state is non-magnetic and a finite energy gap exists in the magnetic energy spectrum. The dimerized singlet ground-state of SrCu$_2$(BO$_3$)$_2$ is separated from the excited triplet states by a gap $\Delta \approx 35$ K [86,87].

As shown in Fig. 15, at zero magnetic field, the thermal conductivity along and perpendicular to the 2D magnetic planes show a double-peak structure obviously [81]. In addition, the high-*T* maximum is field-independent and the low-*T* maximum is suppressed strongly by magnetic field. The magnetic contribution to the thermal conductivity was not evidenced, since the double-peak structure is isotropic and the magnetic excitations are localized. The authors' quantitative analysis in terms of resonant scattering of phonons by magnetic excitations explained the double-peak structures and the magnetic field dependence and gave evidence for strong spin-phonon coupling [81].

Recently, the similar behavior was found in the thermal conductivity of the diamond-chain compound Cu$_3$(CO$_3$)$_2$(OH)$_2$ (CHC) [88]. CHC is the first experimental realization of 1D distorted diamond-chain model [89], in which the dimerized singlet ground state separated from the excited triplet states by a gap $\Delta \approx 50$ K [88]. The earlier studies indicated that CHC can be simply described as a combination of spin-dimer and spin-chain systems. Recently, an effective generalized spin-1/2 diamond chain model [90], with a dominant next-nearest-neighbor AFM dimer coupling $J_2$, two AFM nearest- and third-nearest-neighbor dimer-monomer exchanges $J_1$ and $J_3$, and a significant direct monomer-monomer exchange $J_m$, was proposed to explain most of the experimental results.

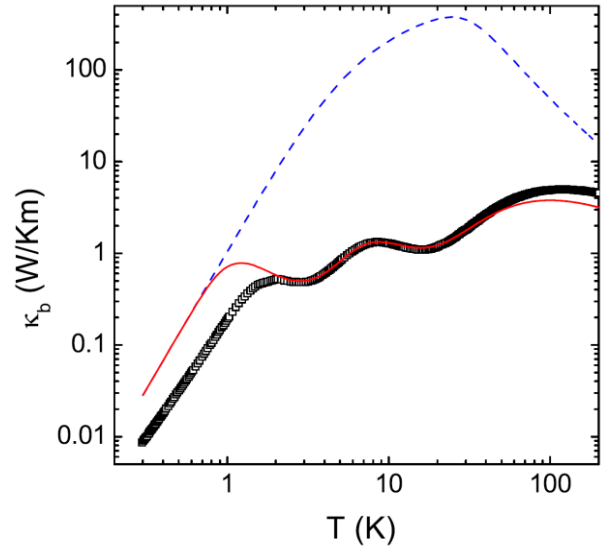

**Figure 16** (Color online) Comparison of the $\kappa(T)$ data of Cu$_3$(CO$_3$)$_2$(OH)$_2$ single crystal and the fitting of Debye model (the solid line) [88]. The dashed line shows the calculated results with switching off the resonant scatterings.

Figure 16 shows the temperature dependence of $\kappa$ of CHC singe crystals [88]. A unique feature of this curve is a three-peak structure at about 2, 8 and 100 K, respectively. The positions of these peaks are quite different from that of phonon peak in insulators, which usually locates at 10–20 K [76]. Due to the small values of $\kappa$, these peaks are actually caused by the presence of two valley-like minimums at 3 K and 17 K. Generally speaking, the possible reasons of these minimums in $\kappa(T)$ curves of the magnetic materials could be either the strong phonon scattering by critical spin fluctuations at some magnetic phase transitions or the phonon resonant scattering processes [43,76,78,81,91-99]. According to the magnetic susceptibility and specific heat results, it is known that CHC does not exhibit any magnetic phase transition at 3 K and 17 K. Therefore, the possible reason is the phonon resonant scattering.

Wu *et al.* tried a more quantitative analysis on the $\kappa(T)$ data by using a classical Debye model for the phonon thermal conductivity [88]. With formula (13), the authors proposed two different resonant scattering mechanisms for explaining the 3 K and 17 K minimums [88,100,101]. The higher-$T$ resonant scattering was attributed to the localized singlet-triplet excitations of spin dimers, with the resonant scattering rate [43,81]

$$\tau_{res1}^{-1} = C \frac{\omega^4}{\left(\omega_1^2 - \omega^2\right)^2} F(T), \tag{18}$$

where $C$ is a free parameter while $F(T)$ describes the difference of thermal populations of the excited triplet and the ground singlet states. It is known that

$$F(T) = 1 - \frac{1 - \exp(-\Delta_1/T)}{1 + 3\exp(-\Delta_1/T)}, \tag{19}$$

with $\Delta_1 = \hbar\omega_1/k_B T$ the energy gap of magnetic spectrum. The lower-$T$ resonant scattering is likely related to the spin-chain excitations, with the resonant scattering rate [93,97,102,103]

$$\tau_{res2}^{-1} = D \frac{\omega^4}{\left(\omega_2^2 - \omega^2\right)^2} \left[1 - \tanh^2 \frac{\Delta_2}{2T}\right], \tag{20}$$

where $\Delta_2 = \hbar\omega_2/k_B T$ is the gap of spin excitations. The total resonant scattering rate is written as

$$\tau_{res}^{-1} = \tau_{res1}^{-1} + \tau_{res2}^{-1}. \tag{21}$$

Figure 16 shows the best fitted result to $\kappa(T)$ with parameters $\Delta_1$ = 50.5 K, and $\Delta_2$ = 7.5 K [88]. They are comparable to the intra-dimer and intra-monomer exchanges (58.0 K and 8.14 K), respectively, determined by the specific heat [88]. In particular, $\Delta_2$ is close to the gap of ~ 1.2 meV at the zone boundary detected by the neutron scattering [104]. Based on this fitting, one can switch off the resonant scatterings by setting $C$ = 0 and $D$ = 0. The phonon thermal conductivity obtained in this way is much larger than the experimental data, as shown in Fig. 16. As expected for resonant scattering, the damping of the phonon heat transport is most pronounced for some particular temperature ranges determined by the magnetic gaps. However, the magnetic scatterings of CHC are actually strong in a very broad temperature range.

It seems that phonon resonant scattering by magnetic excitations is very common in spin-dimer systems which have a gap between the singlet ground state and the excited triplet state.

### 3.5 Heat transport in the Spin-Peierls system

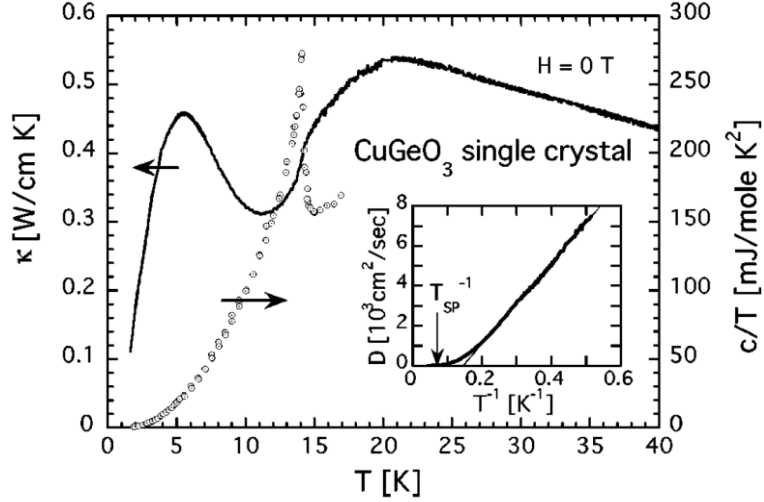

**Figure 17** Thermal conductivity and specific heat of CuGeO$_3$ single crystal in $H = 0$ [105]. Inset: Low-temperatures diffusivity extracted by dividing k by the fitted lattice specific heat.

The thermal conductivity of SP systems was firstly studied by Ando *et al.* [105] In this SP material, the size of the spin gap has been found to be $\Delta \approx 23$ K [18]. This gap can be suppressed with magnetic fields; above a critical field $H_c \approx 12.5$ T, the system undergoes a first-order transition to an incommensurate phase [106]. Figure 17 shows the temperature dependence of thermal conductivity of CuGeO$_3$ single crystal in the temperature range 1.5–40 K. Also shown in this figure are the specific heat data measured on the same crystal in the temperature range 1.8–16 K [105]. The peak of the specific heat is associated with the SP transition [107]. With decreasing temperature from 40 to 1.5 K, the $\kappa$ increases and peaks at ~ 22 K. Below this temperature, the curves of $\kappa$ starts to drop rapidly. At the SP transition temperature (14 K), $\kappa(T)$ shows a kink towards a faster drop. With decreasing temperature, the second peak occurs at about 5.5 K.

Note that the first peak of $\kappa(T)$ appears at a lower temperature than the susceptibility (typically 55 K [20]). At higher temperature region, the phonon thermal conductivity $\kappa_{ph}$ approach to be constant. A constant phonon background about 0.4 W/cm K can be subtracted in the temperature range 15 to 50 K and the remaining contribution approximates $\kappa_s$ which has a peak at about 22 K. Below the SP transition temperature, it is reasonable to assume that $\kappa$ is only due to phonons since $\kappa_s$ diminishes [105]. The phonons are scattered by both spin excitations and crystal defects. Hence, the second peak at about 5.5 K can be formed in the thermal conductivity curve.

Figure 18 shows the $\kappa(T)$ curve in the field range 0–16 T [105]. With increasing magnetic field, one remarkable feature in this curve is the strong suppression of the 5.5 K peak. The strong suppression of the 5.5 K peak with magnetic field is related to the magnetic excitations in the SP phase. This behavior indicated that there exists a strong phonon-magnon scattering in this system. The position of this peak does not shift much below 12 T, indicating that the gap does not change much. However, this peak becomes a shoulder at 14 T and it shows a new upturn below 7 K at 16 T. Meanwhile, the position of the 22 K peak is almost unchanged with the change of magnetic field.

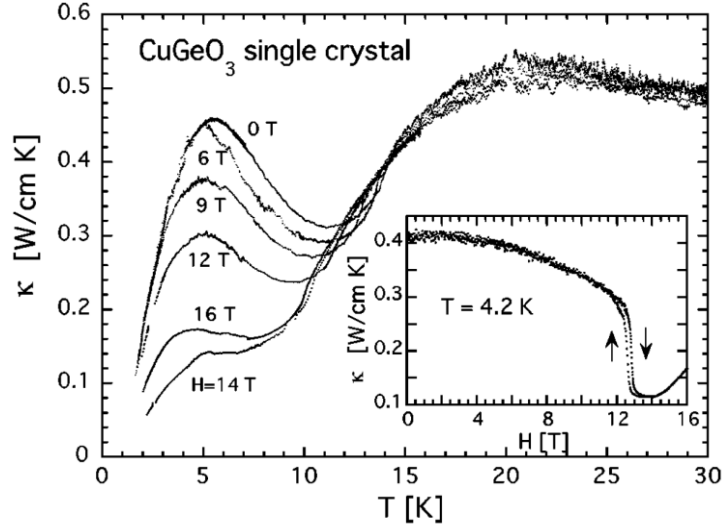

**Figure 18** A set of $\kappa(T)$ of CuGeO$_3$ single crystal in magnetic fields up to 16 T. Inset: $\kappa$ as a function of magnetic fields at 4.2 K [105].

Since the thermal conductivity at low temperatures is greatly suppressed and the diffusivity still has the $1/T$ dependence, the authors concluded that phonons are strongly scattered by solitons [105]. The solitons are the result of magnetic and structural excitations. They are the domain walls in the dimerized lattice and the spin value of solitons is 1/2. Below 10 K, there is an increase of $\kappa$ from 14 to 16 T. In this temperature region, the magnetic specific heat is proportional to $T^3$ due to the gapless phason modes [108]. The increase of phason excitations with increasing field can explain the increase of $\kappa$. Another possibility is that the solitons themselves have the ability to carry heat.

Impurity-substitution effect in CuGeO$_3$ has been studied by Takeya et al. [109] In CuGeO$_3$, a small amount of impurity leads to an exotic low-$T$ phase, namely, dimerized antiferromagnetic phase which can be understood as a state of spatially modulated staggered moments accompanied with the lattice distortion [110,111]. Moreover, when the impurity concentration $x$ exceeds a critical concentration $x_c$, a uniform AFM phase appears below the Néel temperature $T_N \sim 4$ K, with the 1D quantum spin-liquid state above the transition. The spin-gap opening suppresses $\kappa_s$ and enhances $\kappa_{ph}$. Examining the $x$ dependence of the spin gap, it turned out that the local spin gap opens at a temperature ($T^*$) independent of $x$, suggesting that the suppression of SP ordering is due to the reduction of the spin diffusion caused by the impurity scattering [109].

## 4 Magnon BEC system

In the magnon BEC system, the behavior of the thermal conductivity is an attractive issue, because very large magnon transport may be expected from the analogy of thermal conductivity in the superfluid (BEC) state of $^4$He [112,113]. However, depends on the different roles of magnons in the heat transport, either an enhancement or suppression of $\kappa$ could be expected. At the temperature significantly smaller than the spin gap of the quantum paramagnets, the magnetic excitations are well populated at the BEC state ($H_{c1} < H < H_{c2}$) because of the gapless dispersion. If the magnons contribute to the heat transport as carriers, the thermal conductivity should be enhanced in the BEC state. On the other hand, the phonon thermal conductivity should be suppressed if magnetic excitations scatter phonons.

Sun *et al.* firstly studied the low-$T$ thermal conductivity of $NiCl_2$-$4SC(NH)_2$ (DTN) [114]. The Ni spins are strongly coupled along the tetragonal $c$ axis, making DTN a system of weakly interacting spin-1 chains with single-ion anisotropy larger than the intra-chain exchange coupling. The anisotropy, intra-chain, and inter-chain exchange parameters of Ni spins were determined to be $D$ = 8.9 K, $J_c$ = 2.2 K, and $J_{a,b}$ = 0.18 K [24,115-117]. The magnon BEC in DTN is a magnetic field (|| $c$) induced AFM ordered state, with the lower and upper critical fields $H_{c1}$ = 2 T and $H_{c2}$ = 12 T [117]. In addition, the maximum of field dependent critical temperature is about 1.2 K, where $H_{c1}$ and $H_{c2}$ merge.

As shown in Fig. 19, the $\kappa(H)$ curves at very low temperatures shows anomalies at two critical transition fields of $H_{c1}$ = 2.5 T and $H_{c2}$ = 12.5 T [114]. Apparently, the main difference between $\kappa_{ab}$ and $\kappa_c$ can only come from anisotropy magnetic contributions to heat transport, acting as either heat carriers or phonon scatterers. Because of the strong anisotropy of the magnon dispersion [24], it is naturally concluded that the strong suppression of $\kappa_{ab}(H)$ at $H_{c1} < H < H_{c2}$ is mainly due to the phonon scattering by magnons; in addition, although the magnon scattering can also weaken the phonon thermal conductivity along the $c$ axis, the magnons with strong dispersion in this direction can act as heat carriers and make an additional contribution to the heat transport. Also shown in Fig. 19 are the $\kappa(H)$ curves with $H \parallel ab$, which cannot induce magnon BEC. It can be seen that at very low temperatures, both the $\kappa_{ab}$ and $\kappa_c$ do not change with field.

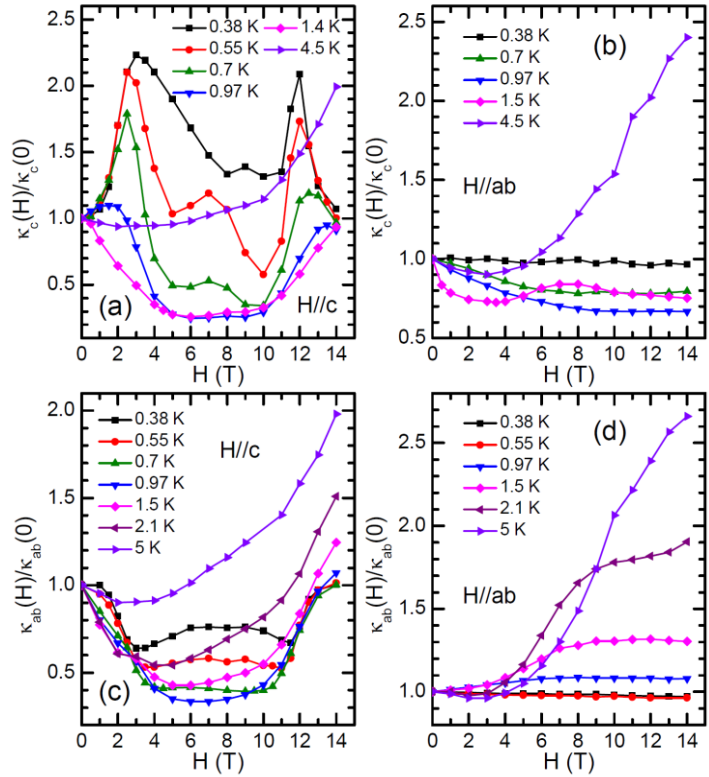

**Figure 19** (Color online) Magnetic-field dependences of thermal conductivities of DTN single crystals at low temperatures [112].

Similar results were later reported by Kohama et al. [118]. Therefore, the magnons could act as heat carriers in DTN, although the present data showed that the magnon heat transport is significant only at critical fields.

In the case of another magnon BEC compound, $Ba_3Mn_2O_8$, Ke et al. found that the magnons in the BEC state are scattering phonons rather than transporting heat [119]. $Ba_3Mn_2O_8$ is an $S = 1$ dimerized quasi-2D antiferromagnet with a gapped ground state, and the excitation gap was estimated as $\Delta = 12.3$ K [120]. Figure 20 shows the magnetic field dependencies of $\kappa_c$ and $\kappa_{ab}$ for both $H \parallel ab$ and $H \parallel c$ [119]. It is clear that they are nearly isotropic on the field direction. The overall behavior of $\kappa(H)$ is that the magnetic field strongly suppresses thermal conductivity. At $T < 0.87$ K that magnon BEC can occur, $\kappa(H)$ show minima at the critical fields. Since the zero-field thermal conductivity of $Ba_3Mn_2O_8$ is purely phononic, the field-induced suppression of $\kappa$ is clearly a result of the scattering effect on phonons by magnetic excitations. The low-field quantum disordered phase has a finite spin gap, which can be weakened by the Zeeman effect. At a fixed temperature, with increasing the magnetic field, the spin gap decreases and the number of low-energy magnons increases quickly, which can strongly scatter phonons and leads to a significant suppression of $\kappa$. It seems that the scattering effect is strongest at the critical fields.

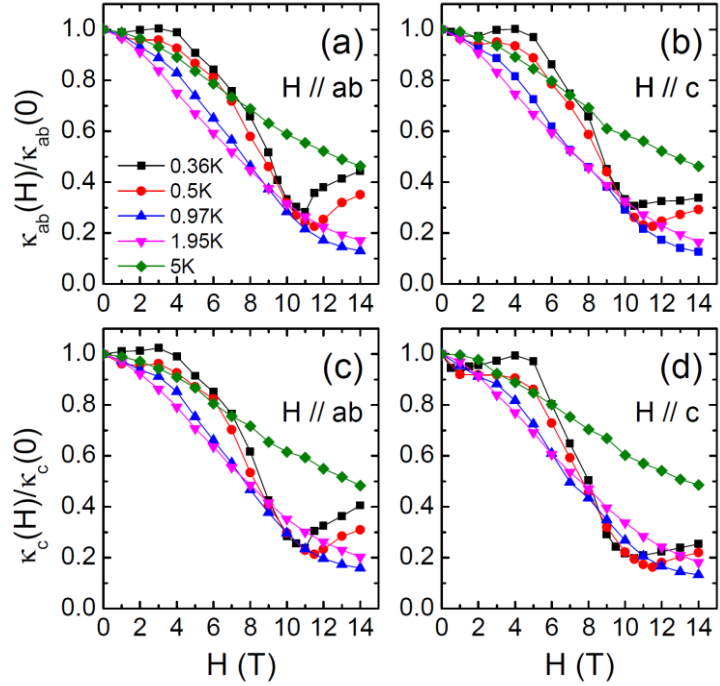

**Figure 20** (Color online) Magnetic-field dependencies of thermal conductivity of $Ba_3Mn_2O_8$ single crystals at low temperatures [119].

Another magnon heat transport phenomenon was recently observed in the $S = 1/2$ spin ladder materials $(CH_3)_2CHNH_3CuCl_3$ (IPA-CuCl$_3$) by Zhao et al. [121] IPA-CuCl$_3$ crystallizes in a triclinic structure and the $Cu^{2+}$ spins ($S = 1/2$) form ladders along the $a$ axis, with rungs along the $c$ axis. The zero-field ground state is quantum paramagnetic with a spin gap of 1.17 meV. When the magnetic field closes the gap at $\mu_0 H_{c1} \approx 10$ T, an AFM state is developed. This has been proposed as a magnon BEC state since the neutron scattering indicated a gapless mode [112,122].

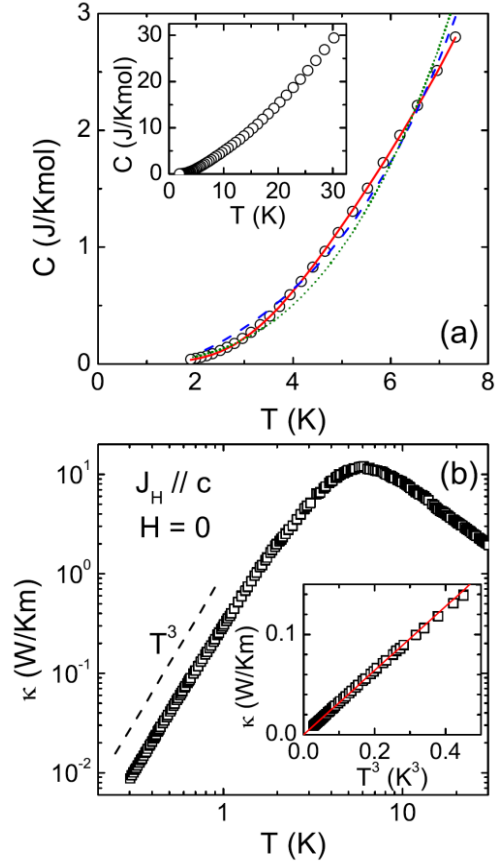

**Figure 21** (Color online) (a) Low-temperature specific heat of an IPA-CuCl$_3$ single crystal in zero field. The inset shows data in a broader temperature range from 2 to 30 K. (b) Temperature dependence of thermal conductivity of an IPA-CuCl$_3$ single crystal in zero field and at 300 mK–30 K. The dashed line indicates $T^3$ temperature dependence. Inset: The low-temperature data in a linear plot for $\kappa$ vs $T^3$. The thin line is a fitting to $\kappa = bT^3$ with the parameter $b = 0.32$ W/K$^4$m for $T < 700$ mK [121].

Figure 21 shows the temperature dependencies of $\kappa$ of an IPA-CuCl$_3$ single crystal in zero magnetic field with the heat current along the $c$ axis [121]. The magnetic excitations can hardly be thermally excited at very low temperatures because of the spin gap (1.17 meV) in the ground state. Therefore, the $\kappa(T)$ is the pure phonon conductivity at low temperatures, which is evidenced by a perfect $T^3$ dependence of $\kappa(T)$ at $T < 700$ mK.

Figure 22 shows the $T$ dependence of the magnon heat transport, the $\kappa_a(T)$ and $\kappa_c(T)$ at $H_{c1}$ of these two samples are measured and compared with their zero-field data [121]. It was found that the $\kappa_a$ ($\kappa_c$) at 9.95 (9.75) T ($H_{c1}$) also follow the $T^3$ dependence at $T < 600$ (700) mK. According to the $\kappa = aT^3$ fittings, one can obtain the value of $a = 2.90$ W/K$^4$m for $\kappa_a$ in 9.95 T and 0.79 W/K$^4$m for $\kappa_c$ in 9.75 T, respectively. The magnon thermal conductivity at $H_{c1}$ can be obtained by subtracting the zero-field data from the critical-field curves, which gives $\kappa_m = 2.08\ T^3$ (W/Km) and 0.29 $T^3$ (W/Km) along the $a$ and $c$ axis, respectively, as shown in Figs. 22(c) and 22(d). This nearly perfect $T^3$ dependence is actually the clearest experimental evidence of the magnon ballistic transport in the AFM systems until now.

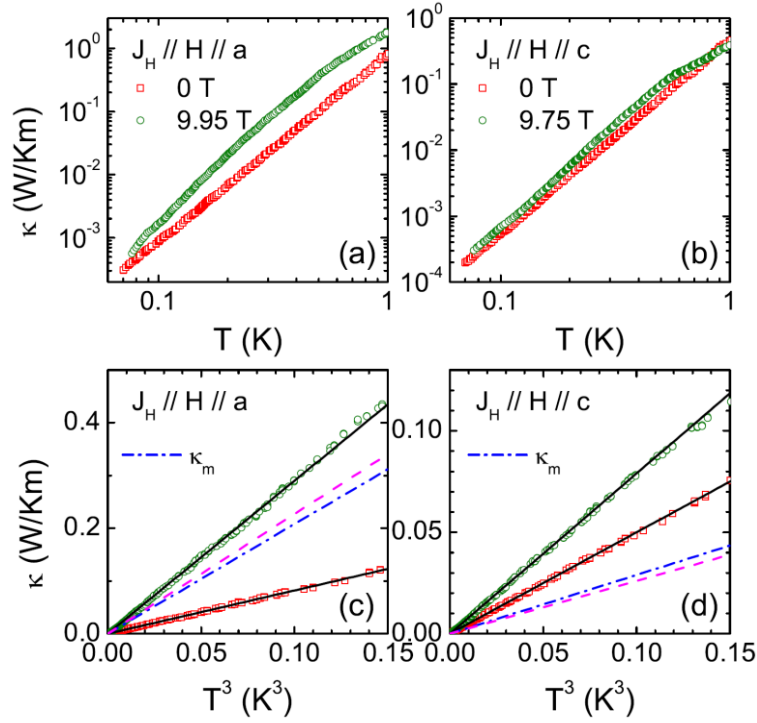

**Figure 22** (Color online) Low-temperature thermal conductivity of IPA-CuCl$_3$ single crystals [121]. (a) Temperature dependencies of the $\kappa_a$ in zero field and 9.95 T ($\parallel a$). (b) Temperature dependencies of the $\kappa_c$ in zero field and 9.75 T ($\parallel c$). [(c) and (d)] Data shown in the $\kappa$ vs $T^3$ plot. The solid lines are linear fittings to the experimental data. The dot-dashed lines denote the magnon thermal conductivity, $\kappa_{m,a} = 2.08\ T^3$ (W/Km) and $\kappa_{m,c} = 0.29\ T^3$ (W/Km), obtained by subtracting the zero-field data from the $\kappa(T)$ curves in the critical fields. The dashed lines show the calculated curves $\kappa_{m,a} = 2.27\ T^3$ (W/Km) and $\kappa_{m,c} = 0.26\ T^3$ (W/Km).

Figure 23 shows the magnetic-field dependencies of $\kappa_a$ and $\kappa_c$ at low temperatures for two IPA-CuCl$_3$ single crystals [121]. The data exhibit two remarkable features. First, the $\kappa$ are field independent at low fields but exhibit a sharp peak at $H_{c1}$, particularly at very low temperatures. It is noted that at $T \to 0$ the peak positions coincide with the transition field of the field-induced AFM state [112,122-125]. Second, the peak feature demonstrates that the gap is closed only at $H_{c1}$ and is reopened above $H_{c1}$, which results in the vanishing of magnon transport in high fields. Therefore, the lowest excitation in the field-induced AFM state is a non-Goldstone mode. In the high-field state with small gap, the magnons can still be easily excited and contribute to transporting heat if $k_B T$ is not smaller than the gap. As shown in Fig. 23, it seems that the $\kappa$ tends to recover its zero-field value at the high-field limit of $H \parallel a$ ($\parallel c$) when $T \le 380$ mK (252 mK) but tends to increase at the high-field limit when $T \ge 520$ mK (380 mK). Therefore, the gap size is estimated to be about 500 mK ($\approx 0.043$ meV) and 300 mK ($\approx 0.026$ meV) for $H \parallel a$ and $c$, respectively. Apparently, such small gaps are beyond the resolution of the earlier neutron measurements. Similarly, TlCuCl$_3$ was the first magnon BEC candidate that showed a gapless Goldstone mode in the field-induced AFM state by the neutron scattering measurement [21] but lately has been found to have a small gap of 0.09 meV by the ESR measurement [126,127].

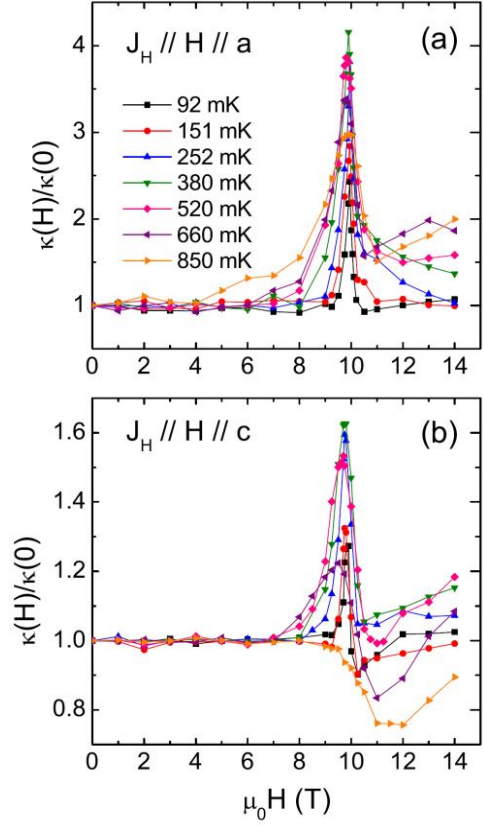

**Figure 23** (Color online) Magnetic-field dependencies of the $\kappa_a$ (a) and $\kappa_c$ (b) of two IPA-CuCl$_3$ crystals at subKelvin temperatures [121].

A characteristic of BEC is the presence of $U(1)$ symmetry, which corresponds to the global rotational symmetry of the bosonic field phase [25]. In field-induced XY-ordered state, the $U(1)$ symmetry spontaneously gets broken and thus a gapless Goldstone mode is acquired. However, the reopening of the gap is clear evidence for a broken uniaxial symmetry of spin Hamiltonian, which rules out a strict description of the magnetic order in terms of BEC [29]. Therefore, the heat transport data indicate that the BEC model has limited applicability to IPA-CuCl$_3$, similarly to TlCuCl$_3$. The theoretical works actually had predicted a general instability of an axially symmetric magnetic condensate toward a violation of this symmetry and the formation of an anisotropy gap at $H > H_{c1}$ [29,128]. It is related to the presence of anisotropic interactions, such as the dipole-dipole coupling, the spin-orbital interaction, etc. [29,128].

## 5 Conclusion

In this article, the heat transport properties of spin-gapped quantum magnets are reviewed mainly from the experimental respect. A large spin thermal conductivity was observed in the two-leg Heisenberg $S = 1/2$ ladder compounds at high temperatures, with the mean free path of spin excitations reaching several thousand Å. The characteristic of magnetic thermal transport of the Haldane chain systems is still an open question. There are controversial results on both the theoretical and experimental results. For the spin-Peierls system, the spin excitations can also act as heat carriers. In spin-dimer compounds, the magnetic excitations mainly play a role of scattering phonons. The magnetic excitations in the so-called magnon BEC systems displayed dual roles, carrying heat or scattering phonons, in different materials. The reason for determining this role is another open question in the field of spin-gapped quantum magnets. For some magnon BEC candidates, the ultra-low-temperature thermal conductivity measurement provides a sensitive technique for probing the spin gap and examining the validity of BEC model.

*We thank the technical assistance from X. Rao, C. M. Xu, J. Shi, J. C. Wu, and H. S. Xu. This work was supported by the National Natural Science Foundation of China (Grant Nos. 11374277, 11574286, 11404316, U1532147), the National Basic Research Program of China (Grant Nos. 2015CB921201, 2016YFA0300103), and the Opening Project of Wuhan National High Magnetic Field Center (Grant No. 2015KF21).*